\documentclass[pre,twocolumn,superscriptaddress]{revtex4}
\pdfoutput=1

\usepackage{amsmath}
\usepackage{graphicx}

\def\Dth{\Delta\theta}
\def\Dp{\Delta p}

\begin{document}

\title{On the effectiveness of mixing in violent relaxation}

\author{Pierre de Buyl}
\affiliation{Center for Nonlinear Phenomena and Complex Systems \\
 Universit{\'e} Libre de Bruxelles, Code Postal 231, Campus Plaine, B-1050 Brussels, Belgium}
\affiliation{Chemical Physics Theory Group, Department of Chemistry, University of Toronto, Toronto, Ontario, M5S 3H6 Canada}
\author{Pierre Gaspard}
\affiliation{Center for Nonlinear Phenomena and Complex Systems \\
 Universit{\'e} Libre de Bruxelles, Code Postal 231, Campus Plaine, B-1050 Brussels, Belgium}

\begin{abstract}
  Relaxation processes in collisionless dynamics lead to peculiar behavior in systems with long-range interactions such as self-gravitating systems, non-neutral plasmas and wave-particle systems. These systems, adequately described by the Vlasov equation, present quasi-stationary states (QSS), i.e. long lasting intermediate stages of the dynamics that occur after a short significant evolution called ``violent relaxation''. The nature of the relaxation, in the absence of collisions, is not yet fully understood. We demonstrate in this article the occurrence of stretching and folding behavior in numerical simulations of the Vlasov equation, providing a plausible relaxation mechanism that brings the system from its initial condition into the QSS regime. Area-preserving discrete-time maps with a mean-field coupling term are found to display a similar behaviour in phase space as the Vlasov system.
\end{abstract}
\date{\today}

\maketitle


\section{Introduction}

The evolution of collisionless systems poses an interesting challenge in kinetic theory. Indeed, from the lack of a collision term in the kinetic equation ruling the evolution of the system arises the need for another relaxation mechanism.
Unusual relaxation properties are found in many physical systems among which self-gravitating systems, non-neutral plasmas and wave-particle interactions for instance. These systems fall into the categories of long-range interacting systems, a domain of physics that is the object of renewed interest~\cite{long-range-02,long-range-07,campa_et_al_phys_rep_2009}.

One phenomenon in particular has been evidenced numerically \cite{yamaguchi_et_al_physica_a_2004} and theoretically \cite{bouchet_dauxois_pre_2005,jain_et_al_relaxation_times_2007}~: In systems with long-range interactions, increasing the number of particles $N$ causes the system to evolve, with a time scale of order $N^0$, towards an intermediate state that is not the one predicted by statistical mechanics and whose lifetime increases as $N^\delta$, with $\delta > 1$. These states are called quasi-stationary states (QSS) and are equilibria of the continuum limit given by the Vlasov equation~\cite{yamaguchi_et_al_physica_a_2004}. QSS have been observed in the Hamiltonian Mean-Field (HMF) model \cite{latora_et_al_prl_1999}, the free-electron laser \cite{barre_et_al_pre_2004} and also in self-gravitating systems \cite{yamaguchi_pre_2008}. Experimental perspectives regarding these QSS have recently been proposed \cite{bachelard_et_al_jstat_2010}.

The fast evolution on a timescale $N^0$ is termed violent relaxation, following the terminology of the studies on 1D self-gravitating systems.
Early works report that the evolution towards a stationary regime is not driven by two-body encounters and question the nature of the initial evolution of gravitating systems~\cite{henon_1964,lecar_iaus_1966}.
In light of these observations, Lynden-Bell presented in 1967 his \emph{statistical mechanics of violent relaxation in stellar systems} \cite{lynden-bell_1967}.  The aim of his theory is to explain the outcome of collisionless dynamics and the unusual energy distribution found in galactic dynamics. Lynden-Bell's (LB) theory is a statistical theory that takes into account, among other properties of the Vlasov equation, its incompressible character in the single-particle phase space; accordingly, an ergodic-type hypothesis on the dynamics allows one to compute, via an entropy maximization, stationary states of the Vlasov equation. LB's theory predicts the magnetization in the HMF model and provides the most likely QSS solution for this model, as of now \cite{antoniazzi_et_al_prl_2007}.
The work of Lynden-Bell has been followed by numerous numerical studies on violent relaxation, see Refs.~\cite{hohl_NASA_1968,luwel_severne_1985,mineau_et_al_numerical_holes_1990,funato_et_al_not_relaxation_1992} for instance.
LB's theory or other attempts to compute the outcome of violent relaxation can be found in more recent work on the self-gravitating sheet model \cite{yamaguchi_pre_2008}, non-neutral plasmas \cite{levin_et_al_plasmas_prl_2008} and 2D self-gravitating systems~\cite{teles_et_al_2d_self-grav_2010}.
Recently, new macroscopic observables have been proposed to measure the approach to equilibrium in the self-gravitating sheet model~\cite{joyce_worrakitpoonpon_jstat_2010}. An extensive study by the same authors indicate the physical situations in which LB's theory provides a good prediction and the reasons it does not work in other situations~\cite{joyce_worrakitpoonpon_pre_2011}.

Let us also mention the use of the Vlasov equation in the field of plasma physics~\cite{balescu_statistical_dynamics} that is probably the widest area of research making use of it. The understanding of collisionless relaxation is also of great interest in this field and is mostly known for the famous problem of Landau damping (see Ref.~\cite{oneil_coroniti_rmp_1999} for instance).

The purpose of this article is to demonstrate via direct numerical simulation of the Vlasov equation that stretching and folding structures occur in some regions of phase space. A measure of the consequent deformation of the fluid allows us to characterize the short-time evolution in the Vlasov equation.
We choose to take as the main vehicle of our study, the Hamiltonian Mean-Field (HMF) model \cite{antoni_ruffo_1995}, a system of globally coupled rotators moving on a circle.
The HMF model has been the object of many studies as a paradigmatic representative of long-range interacting systems and the properties of its associated Vlasov dynamics are well-known. In addition, in the spirit of demonstrating fundamental dynamical properties, the similarity of the HMF model with the forced pendulum makes it a worthwhile example, as evidenced in Ref.~\cite{elskens_escande_nonlinearity_1991}.

Besides, we also consider discrete-time versions of Vlasov dynamics, in which the phase-space distribution function evolves under the effect of an area-preserving map.  Two such mean-field maps are investigated, which are based on Arnold's cat map \cite{PV87,S02} and the standard map \cite{C79,LL83}.

The plan of the paper is the following.
In Section~\ref{sec:HMF}, we introduce the Hamiltonian Mean-Field model, its associated Vlasov equation and the phenomenology of QSS.
Section~\ref{sec:perim} presents the numerical algorithm and the computation of the perimeter that is used to substantiate quantitatively our claims.
Section~\ref{sec:stretch} gives the results of Vlasov simulations, in which stretching and folding occur in phase space. In Section~\ref{sec:mfm}, mean-field maps are studied in order to determine the conditions of stretching of fluid boundaries in phase space.

\section{Collisionless dynamics and the Hamiltonian Mean-Field model}
\label{sec:HMF}

The Hamiltonian Mean-Field (HMF) model was introduced as a simplified model to study collective effects and has become a paradigmatic model to study long-range interactions \cite{antoni_ruffo_1995,campa_et_al_phys_rep_2009}. It is composed of particles on a circle interacting via a cosine potential and described by the following Hamiltonian~:
\begin{equation}
  \label{eq:HMF}
  H = \sum_{i=1}^N \frac{p_i^2}{2} + \frac{1}{2N} \sum_{i,j=1}^N \left(1-\cos(\theta_i-\theta_j)\right)\; ,
\end{equation}
where $\theta_i$ is the position in the interval $[-\pi;\pi[$ of the $i$th particle, $p_i$ its momentum and $N$ is the number of particles.
At equilibrium, the HMF model is characterized by a second-order phase transition, identified by the magnetization~:
\begin{equation}
  {\bf m}=m_x + i \, m_y = \frac{1}{N} \sum_{j=1}^N e^{i\theta_j} ~.
\end{equation}
$m=|{\bf m}|$ is equal to zero above the critical energy $u_c = \frac{H}{N} = \frac{3}{4}$, while $m > 0$ for energies below $u_c$.

Starting from an out-of-equilibrium initial condition, the HMF model displays interesting dynamical phenomena~: in addition to the energy, the initial value of $m$ influences the regimes attained by the system \cite{antoniazzi_et_al_prl_2007}, a dependence which is not found at equilibrium.
The value that $m$ reaches is not the one corresponding to equilibrium statistical mechanics. This phenomenon lasts for a time lapse that depends on the number $N$ of particles considered as $N^{1.7}$ after which the system eventually relaxes to equilibrium. This intermediate regime is called a QSS~\cite{latora_et_al_prl_1999,barre_et_al_pre_2004}.

Let us now introduce the Vlasov equation describing the evolution of the distribution function $f(\theta,p)$ in the HMF model and valid in the thermodynamic limit~\cite{balescu_statistical_dynamics,antoniazzi_califano_prl}~:
\begin{eqnarray}
  \label{eq:vlasov}
  \frac{\partial f}{\partial t} &+& p\, \frac{\partial f}{\partial \theta} - \frac{dV[f]}{d\theta} \frac{\partial f}{\partial p} = 0\; , \cr
    & & \cr
  V[f](\theta) &=& 1 - m_x[f] \cos\theta - m_y[f] \sin\theta\; , \cr
  m_x[f] &=& \int d\theta\, dp\, f \cos\theta \; ,\cr
  m_y[f] &=& \int d\theta\, dp\, f \sin\theta\; ,
\end{eqnarray}
where $\theta$ is the periodic spatial coordinate, $p$ is the momentum, $V$ is the potential, depending self-consistently on $f$.
The time evolution of Eqs.~(\ref{eq:vlasov}) conserves the energy $U$~:
\begin{align}
  \label{eq:U}
  U[f](t) = \int d\theta\ dp\  f(\theta,p ; t) & \quad \cr
  \times \left( \frac{p^2}{2} + \frac{1}{2} \left( 1 - m_x[f]\cos\theta - m_y[f]\sin\theta \right) \right) ~,
\end{align}
the normalization $\int d\theta\, dp\,  f(\theta,p ; t) = 1$ and the total momentum $\int d\theta\, dp\,  f(\theta,p ; t)\, p $.

In order to solve numerically Eqs.~(\ref{eq:vlasov}), we use the semi-Lagrangian method (see Ref.~\cite{sonnendrucker_et_al_semi-lag_1999}, or Ref.~\cite{de_buyl_cnsns_2010} for an application to the HMF model) with cubic spline interpolation. Although numerical limitations come into play \cite{califano_galeotti_pop_2006}, Vlasov simulations have proven useful to study the HMF model in the thermodynamic limit \cite{antoniazzi_califano_prl}.
The semi-Lagrangian method displays a relatively small amount of numerical dissipation and thus fits adequately the purpose of computing the perimeter of the fluid.

Lynden-Bell's (LB) theory has proven so far successful for the HMF model \cite{antoniazzi_et_al_prl_2007}, the free electron laser \cite{barre_et_al_pre_2004} and partly for the self-gravitating sheet model \cite{yamaguchi_pre_2008,joyce_worrakitpoonpon_pre_2011}.
The effectiveness of Lynden-Bell's theory to describe the HMF model has been an important confirmation of the adequateness of the Vlasov equation to understand QSSs.

It has been recently shown that Lynden-Bell's theory captures fundamental properties of the Vlasov dynamics and that phase space \emph{stirring} already provides a fair amount of evolution. The authors of Ref.~\cite{de_buyl_et_al_rsta_2011} obtain good comparison between Lynden-Bell's theory and a model with no interaction at all between particles. The same authors propose an exact solution to the mean-field equilibria in the Vlasov equation, fully demonstrating how stirring allows the dynamics to reach steady states in Vlasov dynamics~\cite{de_buyl_et_al_in_prep}.
Stirring is however insufficient to obtain the better agreement found in the HMF model when using LB's theory. To complete the dynamical picture, another relaxation mechanism is needed.

\section{Evolution of the perimeter of the fluid}
\label{sec:perim}

In the context of 2D fluid dynamics, experimental results can be obtained by direct visual inspection. In the kinetic context, the distribution function (DF) lies in phase space $(\theta,p)$. The technique that we propose is reminiscent of methods used in fluid dynamics. To analyze the complex behavior of the DF, we track its boundary in the same way as tracking a dyed region in a fluid. Instead of a dye, phase space is filled with a step profile of value $f_0$. Inside the step, $f=f_0$ while outside of the step $f=0$. An example of such step profiles is known as the waterbag (WB) initial condition. The WB initial condition possesses the following advantages~: a simple formulation of LB's theory, a nice interpretation in terms of a dye in phase space and a very widespread use in the literature.

The perimeter $P_f$ of the DF $f$ is the length of the interface between the $f=f_0$ and $f=0$ regions of phase space, it is similar to the {\it intermaterial area per unit volume} defined in Ref.~\cite{ottino_book_1989}.
$P_f$ provides a direct measure of the deformation of the DF.

The WB is defined by a width of $2\Dth$ and $2\Dp$ in the two dimensions of phase space. The initial perimeter $P_f$ of this ``fluid'' is $4 ( \Dth + \Dp )$. The preservation of phase-space volume in Vlasov dynamics implies that the area of the fluid remains constant.

In the simple case of free-streaming, it is easy to compute the time evolution of $P_f$~:
\begin{equation}
  \label{eq:free-s}
  P_f(t) = 4 \Dth + 4 \Dp \sqrt{1+t^2} ~,
\end{equation}
which is seen to be asymptotically linear in time. This linear behavior is expected to hold in the case of regular dynamics, as in the case of integrable systems.

In more complex situations, we expect a deformation of the contour of the fluid, similar to what happens in fluid dynamics \cite{ottino_book_1989}.
Indeed, 2D flows submitted to parametric forcing exhibit chaotic behavior, and analogies between Vlasov dynamics and 2D Euler equations are common \cite{del-castillo-negrete_firpo_chaos_2002}.
If neighboring fluid elements experience an exponential separation of their positions, we expect that the perimeter is dominated by this contribution so that
\begin{equation}
  P_f(t) \propto e^{\lambda t}
\end{equation}
where $\lambda$ is a positive real number.
After the stretching mechanism has taken place, the fluid folds onto itself to accommodate for the increasing perimeter at fixed area. In fluid dynamics, the obtention of an optimal mixing, via stretching and folding of the fluid, is a ongoing topic of interest \cite{ottino_book_1989,wiggins_ottino_rsta_2004}.

\begin{figure}[h!]
  \centering
  \includegraphics[width=\linewidth]{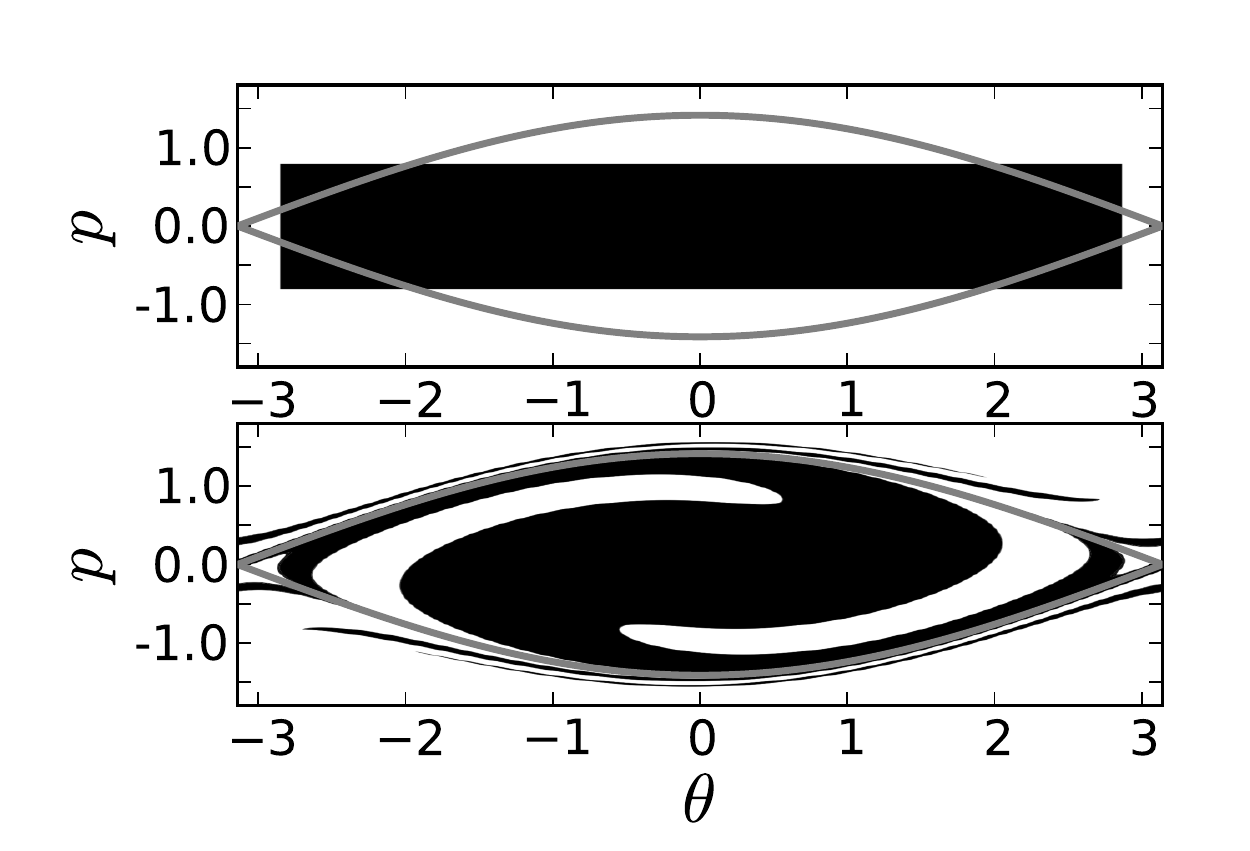}
  \caption{Vlasov simulation for a non-interacting set of pendula. The top panel displays the waterbag initial condition, parametrized by $\Dth=2.85$ and $\Dp=0.79$. The bottom panel displays only the contour at time $t=10$. The filamentary structure is caused, in this system, only by the differential rotation in phase space, i.e., phase-space \emph{stirring}.}
  \label{fig:pen}
\end{figure}

\begin{figure}[h!]
  \centering
  \includegraphics[width=\linewidth]{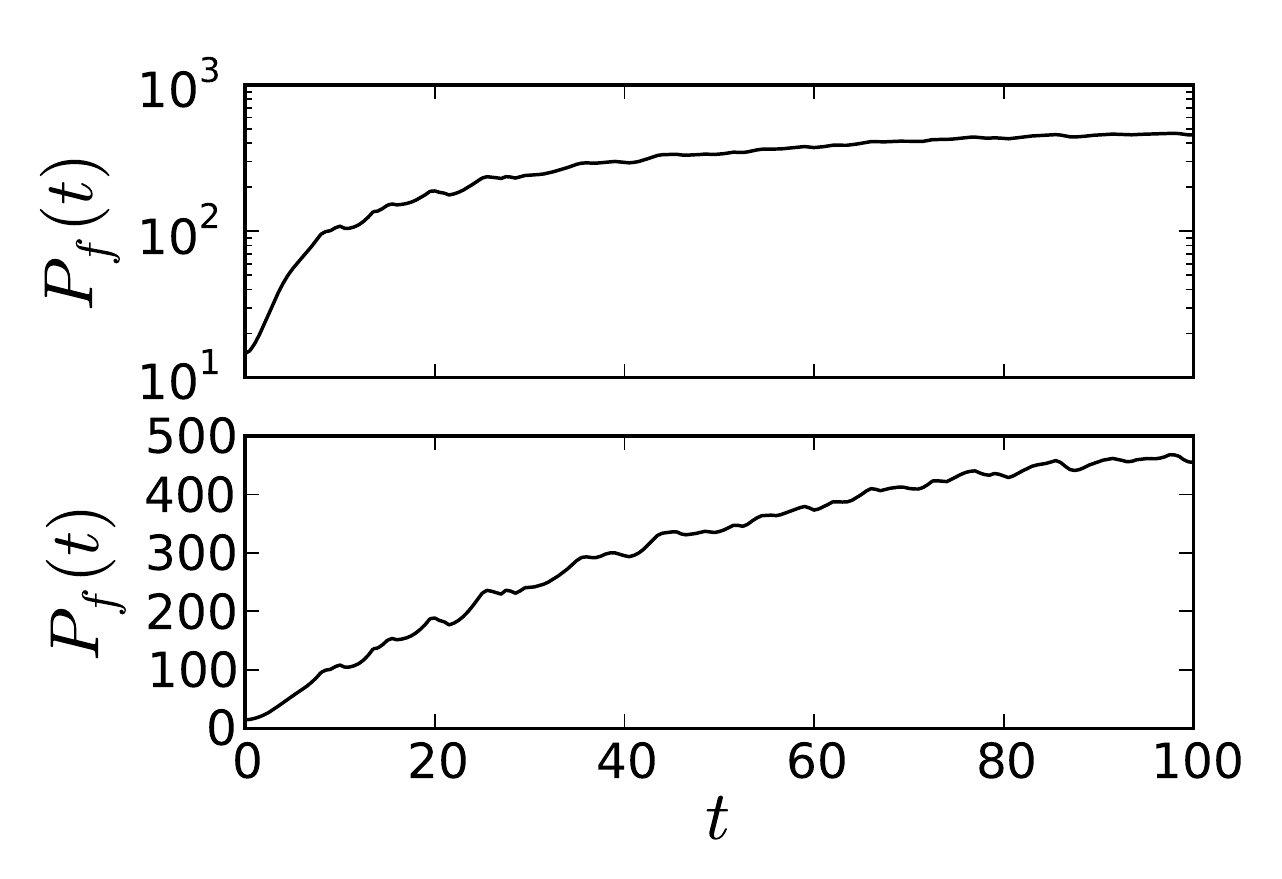}
  \caption{Perimeter of the fluid as a function of time for the simulation shown in Fig.~\ref{fig:pen}. The top panel uses a logarithmic scale on the $y$-axis and the bottom panel uses a linear scale for the $y$-axis.}
  \label{fig:Poft-stir-2k}
\end{figure}

The perimeter is computed numerically with the help of the routine CONREC~\cite{bourke_conrec_1987}. This routine, initially used to display contour lines on a computer display, has been modified in order to compute the length of this contour and has been implemented within the Vlasov simulation program used in Ref.~\cite{de_buyl_cnsns_2010}, a modification that is necessary in order to avoid storing all time steps of the DF for post-processing. Equation~(\ref{eq:free-s}) is checked to give a perfect match with the simulation for $m=0$ (data not shown). A check of our conjecture for non-interacting systems is tested for a pendulum (a system similar to Eqs.~(\ref{eq:vlasov}) but in which the magnetization $\bf m$ is set to a constant value in the equations of motion). Figure~\ref{fig:pen} displays the evolution of the DF and Fig.~\ref{fig:Poft-stir-2k} displays the evolution of $P_f(t)$ for the same simulation.

Table~\ref{tab:params} lists the numerical parameters for the Vlasov simulations. All initial conditions are runs with all these parameters, allowing to monitor the convergence of the physical quantities as a function of the number of grid points.
\begin{table}[!ht]
\caption{\label{tab:params}Numerical parameters for the Vlasov simulations.}
\begin{center}
\begin{tabular}{c @{\hskip 1cm } l l l }
\hline
 & $N_\theta$ & $N_p$ & $\Delta t$ \\
\hline\hline
0k5 & 512  & 512 & $0.1$ \\
1k & 1024  & 1024 & $0.1$ \\
2k & 2048  & 2048 & $0.1$ \\
4k & 4096  & 4096 & $0.1$ \\
8k & 8192  & 8192 & $0.1$ \\
8kb & 8192  & 16384 & $0.1$ \\
\hline
\end{tabular}
\end{center}
\end{table}

\section{Stretching and folding in phase space}
\label{sec:stretch}

The present section focuses on the interacting HMF model.
The variations of the self-consistent field ${\bf m}$ cause a complex evolution of the initial waterbag that we consider in several test-cases.
The first case (run f in Table~\ref{tab:simus}) leads to a magnetized state and has the property that $m_x$ keeps its sign in the course of time. Also, a significant amount of mass remains enclosed inside the separatrix at all times, meaning that it is not subject to alternating stretching motion but only to a stirring motion that is reminiscent of the non-interacting situation.
A variation on run f is the run j that also leads to a magnetized regime and possesses a similar structure in phase space. $m_0$ for run j is however equal to the predicted value of $m$ in the QSS regime by Lynden-Bell's theory and we expect a smaller amount of stretching and folding because the variations of $m$ are smaller.
The second case (run i in Table~\ref{tab:simus}) refers to the peculiar value $U=0.69$, with $m_0=0.20$, which is known to possess counter rotating resonances. $m_x$ then experiences changes of sign that imply elliptic-hyperbolic bifurcations~\cite{del-castillo-negrete_firpo_chaos_2002} which we discuss further on in section~\ref{sec:ell-hyp}.

\begin{table}[!ht]
\caption{\label{tab:simus}Parameters for the waterbag initial condition of runs f, i and j.}
\begin{center}
\begin{tabular}{c @{\hskip 1cm } l l}
\hline
Run & $U$ & $m_0$\\
\hline\hline
f & 0.60  & 0.20\\
i & 0.69  & 0.20\\
j & 0.565 & 0.50\\
\hline
\end{tabular}
\end{center}
\end{table}

The organization of phase space is similar to the one of a pendulum at every given time.
The simplest situation is when $m_y=0$ at all times with small variations of $m_x$ around a finite average value. The system is then close to a set of pendula, although small oscillations of $m_x$ are sufficient to induce stretching and folding.
In the presence of stronger oscillations of $m_x$, the observed phenomenon of stretching and folding is increased. The oscillations of the separatrix, which drive the stretching and folding, are indeed linked to the one of $m_x$.
The region of phase space that is located under the separatrix at all times experiences deformations, but the organization of phase space is such that around the elliptic fixed point at $(0,0)$ a part of the waterbag remains intact, reminding the ``core-halo'' phenomenon observed in self-gravitating systems~\cite{yamaguchi_pre_2008,joyce_worrakitpoonpon_pre_2011}.

In the counter propagating resonances regime, first observed in Ref.~\cite{antoni_ruffo_1995}, an example of hyperbolic-elliptic bifurcation is found. Such a behavior has been observed already in the single-wave model in Ref.~\cite{del-castillo-negrete_firpo_chaos_2002}. The authors of the latter reference find that a succession of elliptic-hyperbolic bifurcation produces strongly chaotic Lagrangian trajectories and destroys dipolar structures, modifying the general behavior of the system. As it stands for the HMF model, two coherent counter-propagating clusters remain present in the system.

\subsection{Variation of $m_x$ at null $m_y$}

\begin{figure}[h!]
  \centering
  \includegraphics[width=\linewidth]{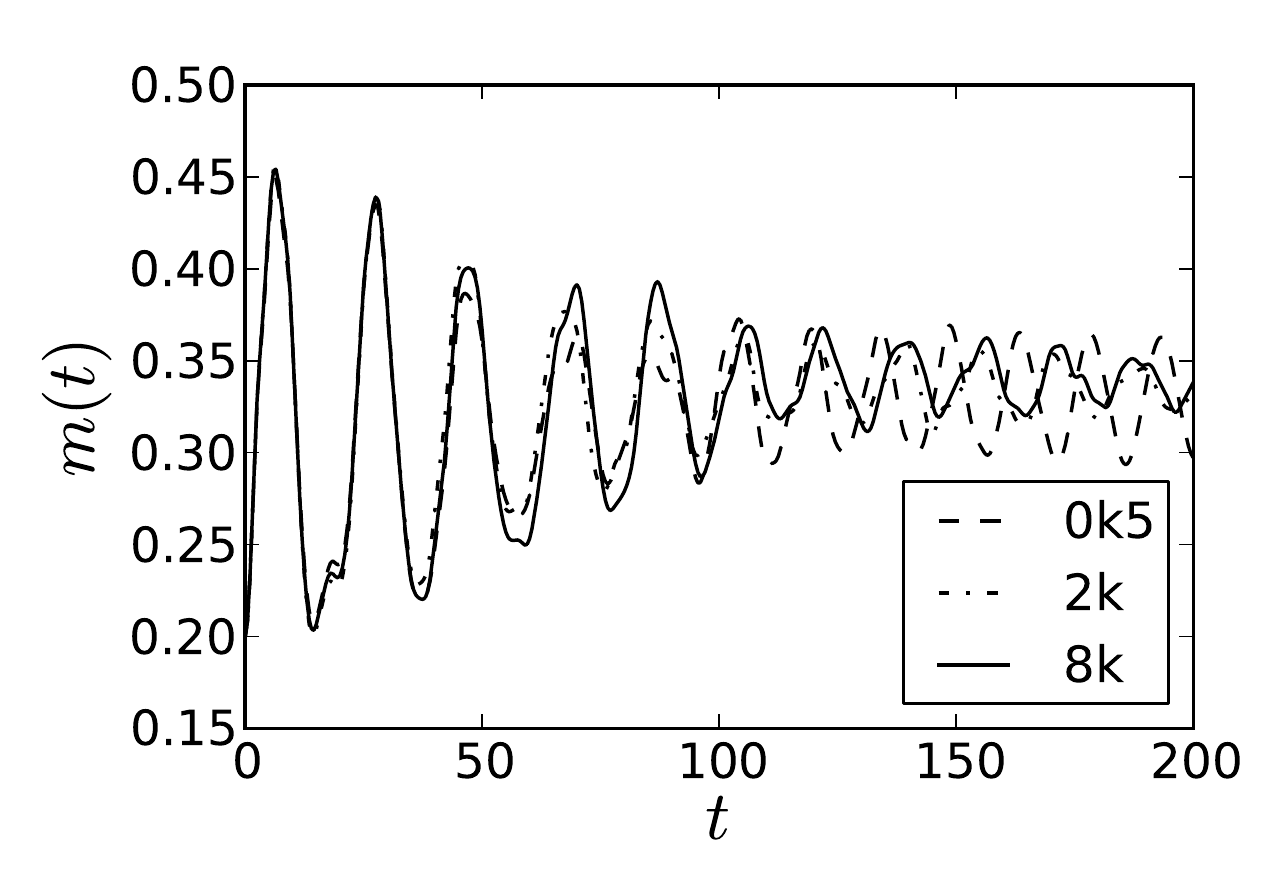}
  \caption{Evolution of $m(t)$ as a function of the time $t$ for run f of Table~\ref{tab:simus} for different numerical settings (see Table~\ref{tab:params}). The initial behavior is very similar for all runs, later times display small phase and amplitude differences.}
  \label{fig:mixf_m(t)}
\end{figure}

We consider run f of Table~\ref{tab:simus}, in this situation, $m_y=0$ at all times and $m_x$ displays strong initial oscillations from $t=0$ to $t\approx 100$ followed by smaller oscillations (see Fig.~\ref{fig:mixf_m(t)}) around a finite value. The initial time lapse corresponds to violent relaxation, during which we are interested in the behavior of the fluid.
The initial stronger oscillations are typical of violent relaxation in self-gravitating systems with the difference that the monitored quantity in these systems is the virial ratio \cite{hohl_NASA_1968}. The value of $m$, in the HMF model, is sufficient to fully follow macroscopic quantities.

Figure~\ref{fig:mixf_m(t)} displays $m(t)$ as a function of time, for run f and the parameters of Table~\ref{tab:params}. Between $t=0$ and $t\approx 50$, the curves superimpose well. Up to $t\approx 100$, qualitative agreement is still valid.
We may already conclude that the amplitude of the oscillations decay, a key point in the theory of violent relaxation.
However, the amplitude of the oscillations in $m(t)$ behave in a different manner than found in Ref.~\cite{de_buyl_cnsns_2010} where increasing the number of grid points enhance sustained oscillations. Here, increasing the number of grid points does not lead to a monotonous increase in the amplitude of the oscillations.
We can quantify this behavior by taking the standard deviation of the time series $m(t)$, taken here between $t=100$ and $t=200$. The results are given in Table~\ref{tab:std}.
The standard deviation is found to decrease from parameters ``0k5'' to ``4k'' and to increase afterward.
This can be understood by the presence of two competing reasons for the damping. A low number of grid points increases the phenomenon of numerical dissipation (as found in Ref.~\cite{de_buyl_cnsns_2010}) while a high number of grid points enables the description of the stretching and folding in phase space. This latter phenomena, by allowing a wider spreading of fluid elements, contributes to the homogeneity of the system, from the point a view of a given energy level, and thus to a reduction of the oscillations in $m(t)$.
\begin{table}[!ht]
\caption{\label{tab:std}Standard deviations of $m(t)$ for run f with different parameters. $m(t)$ is taken between $t=100$ and $t=200$.}
\begin{center}
\begin{tabular}{c c c c c c}
\hline
0k5 & 1k & 2k & 4k & 8k & 8kb\\
\hline\hline
$2.5~10^{-2}$ & $1.7~10^{-2}$ & $1.3~10^{-2}$ & $1.1~10^{-2}$ & $1.5~10^{-2}$ & $1.6~10^{-2}$\\
\hline
\end{tabular}
\end{center}
\end{table}

\begin{figure}[h!]
  \centering
  \includegraphics[width=\linewidth]{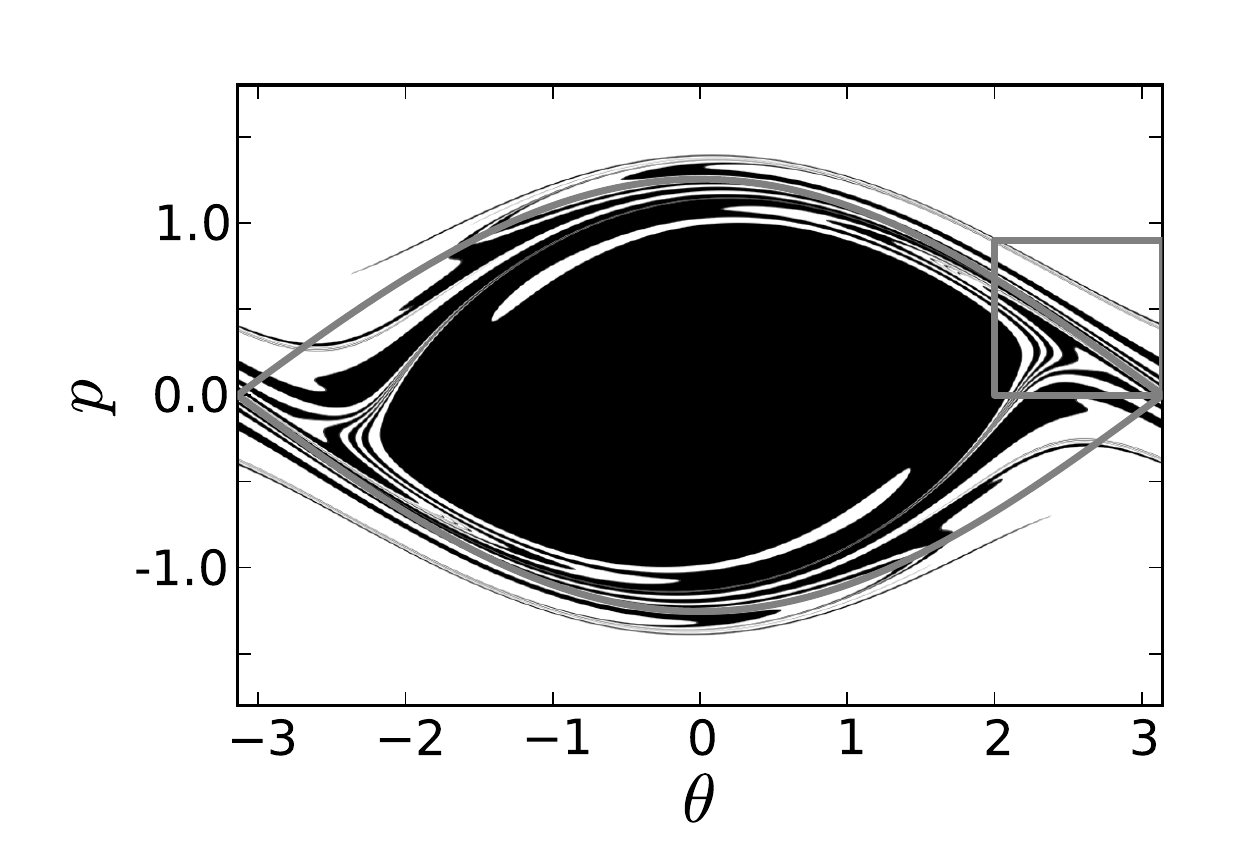}
  \caption{Phase space for run f at time $t=25$, with the setting ``8k'', the contour is displayed, the region in red corresponds to $f=f_0$ and the region in white corresponds to $f=0$. The instantaneous separatrix is drawn and three separate regions are visible~: well inside the separatrix, around the separatrix and well above the separatrix. The filamentary structure is the result of the self-consistent evolution of the system. The box indicates the zoomed region of Fig.~\ref{fig:zoom_mixf}.}
  \label{fig:mixf_contour}
\end{figure}

A general view of phase space is given in Fig.~\ref{fig:mixf_contour}. The phase-space structure is as follows~: there is an elliptic fixed point at $(\theta,p) = (0,0)$ and an hyperbolic fixed point at $(\theta,p)=(\pm\pi, 0)$. A separatrix joins the two hyperbolic points, but its height is variable as the magnetization evolves in time. This time dependence allows one to consider the HMF model as a time-dependent pendulum, keeping in mind that in the HMF model the particles generate self-consistently the value of $\bf m$.
Three different behaviors are found, depending on the energy with respect to the separatrix, consistent with the findings of the forced pendulum~\cite{elskens_escande_nonlinearity_1991}~:
\begin{itemize}
\item Below the separatrix energy, the so-called trapped particles remain close to the elliptical point and perform an oscillatory motion.
\item Above the separatrix, particles cross space with a velocity of constant sign.
\item At energies near the separatrix, particles experience a very complex behavior, known as separatrix crossing.
\end{itemize}
The oscillations of the separatrix will cause fluid elements to experience stretching and folding.

\begin{figure}[h!]
  \centering
  \includegraphics[width=\linewidth]{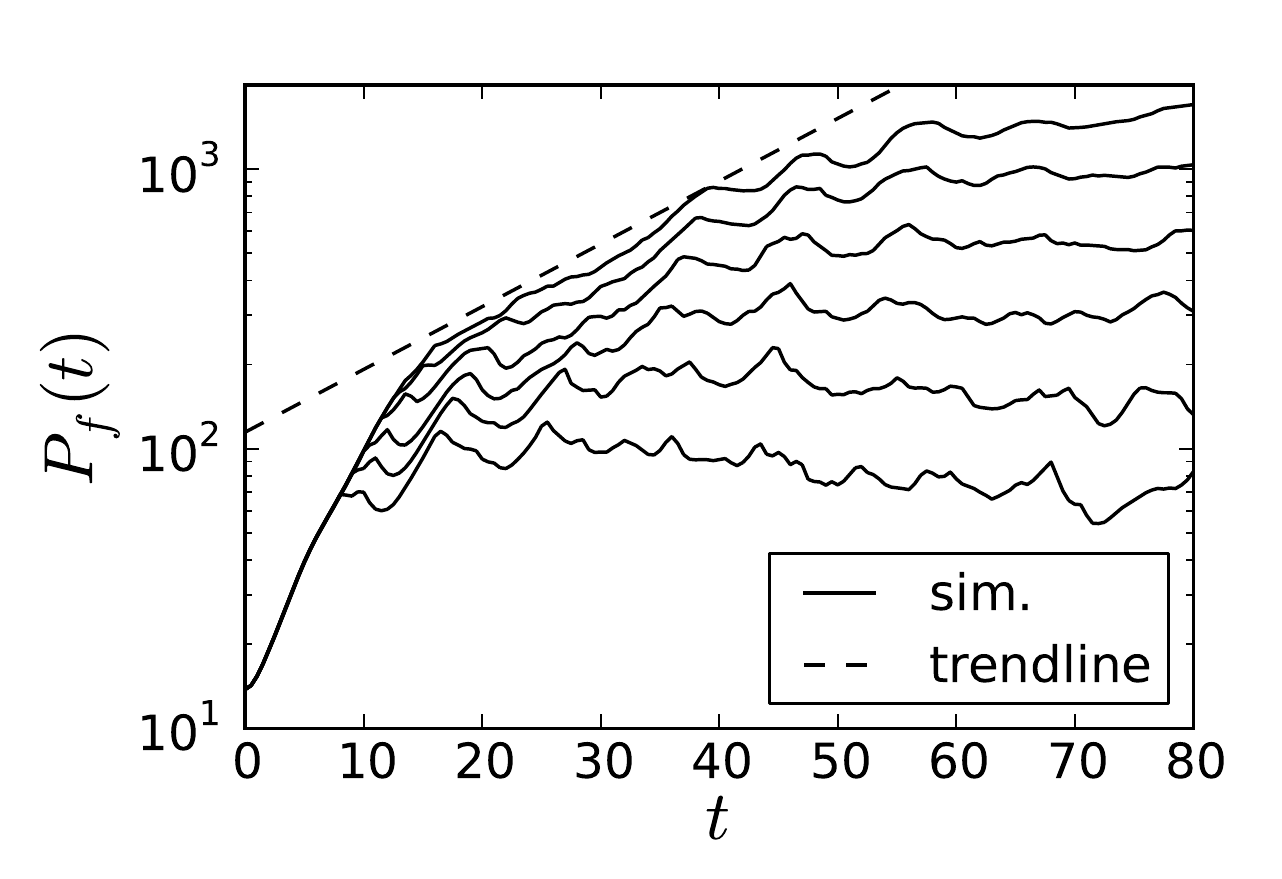}
  \caption{The perimeter $P_f(t)$ for run f, for several configurations of Table~\ref{tab:params}. The initial behavior is an exponential increase of $P_f$ as a function of time (notice the logarithmic scale for the $y$-axis), followed by a saturation due to a numerical limitation. It is observed that increasing the number of grid points increases the saturation level.}
  \label{fig:mixf_perim}
\end{figure}

Figure~\ref{fig:mixf_perim} displays $P_f(t)$ for the parameters from Table~\ref{tab:params}. Increasing the number of grid points allow to reach a higher value of $P_f$ before saturation. Before saturation, all runs show quantitative agreement on the value of $P_f(t)$, indicating convergence as the number of grid points is increased.
The initial behavior of $P_f(t)$ is exponential (notice the logarithmic scale of $P_f$ in Fig.~\ref{fig:mixf_perim}), showing that part of the perimeter is experiencing an early exponential separation.

We notice that the stretching rate $\lambda$ of the perimeter differs from a Lyapunov exponent in several respects.  The perimeter is associated with a certain distribution function so that its stretching rate may depend on the choice of the initial distribution function.  The perimeter may undergo a regime of exponential growth, which does not go on, so that the stretching rate may not be defined in the long-time limit.  Nevertheless, the concept of Lyapunov exponent characterizing chaotic dynamics is useful to interpret the present stretching rate. In Fig.\,\ref{fig:mixf_perim}, the early exponential increases observed for $P_f$ can be understood in terms of the phase-space dynamics of the pendulum periodically driven by the large early oscillations of the mean field seen in Fig.\,\ref{fig:mixf_m(t)}.  Indeed, the periodically driven pendulum is known to be chaotic, which can induce the stretching of phase-space domains leading to a positive Lyapunov exponent.  Therefore, the perimeter can undergo an exponential growth as long as this induced stretching goes on.  However, as time increases, the oscillations of the mean field become of lower amplitudes as observed in Fig.\,\ref{fig:mixf_m(t)}, which reduces the extension of the chaotic zones in the phase space of the periodically driven pendulum and tends to decelerate the growth of $P_f$ as seen in Fig.\,\ref{fig:mixf_perim}.

\begin{figure}[h]
  \centering
  \includegraphics[width=\linewidth]{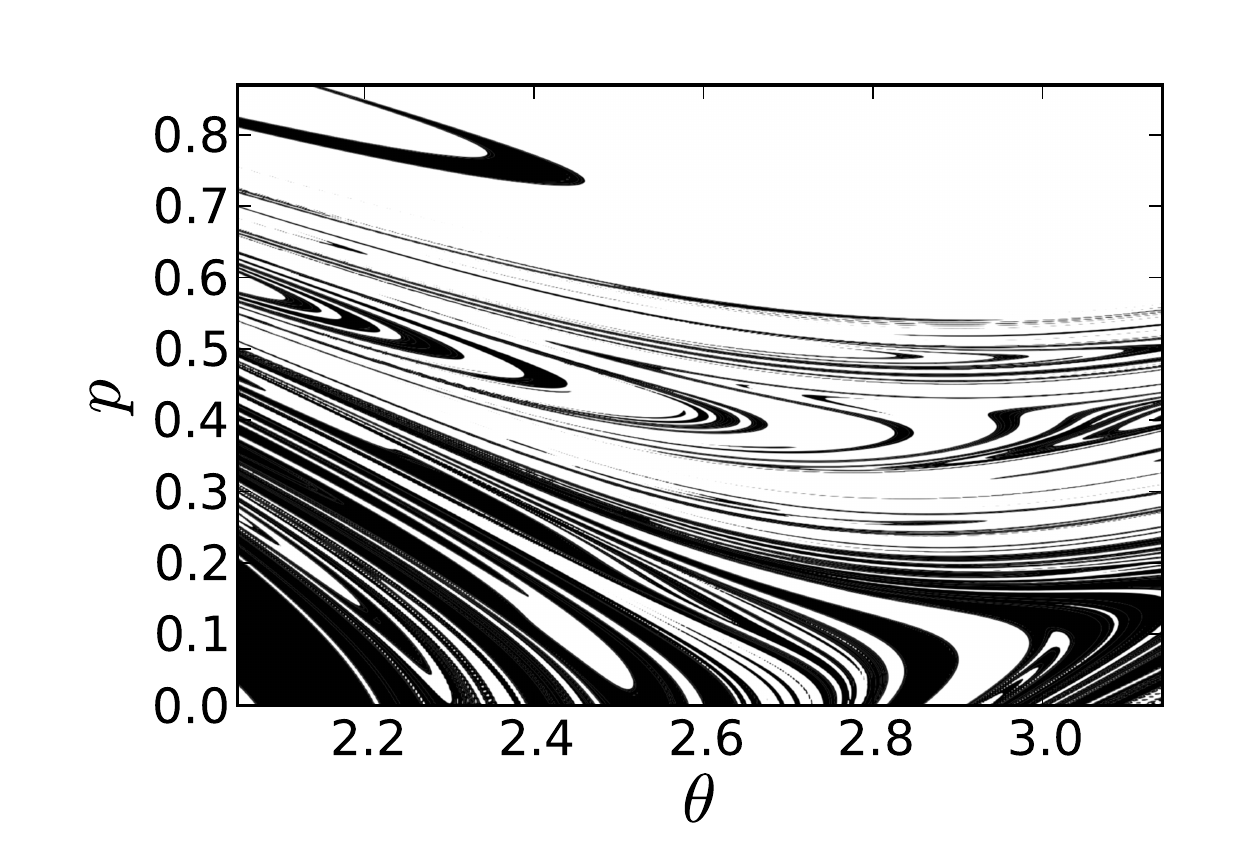}
  \caption{Zoom on Fig.~\ref{fig:mixf_contour} at a later time $t=100$. The complex evolution of the initial waterbag leads to a stretching and folding behavior. A very intricate sequence of $f=0$ and $f=f_0$ regions can be observed.}
  \label{fig:zoom_mixf}
\end{figure}

Finally, to illustrate the behavior at small scales, we display in Fig.~\ref{fig:zoom_mixf} a zoom of Fig.~\ref{fig:mixf_contour} at a later time. The stretching and folding structures are clearly apparent.

\subsection{Elliptic-hyperbolic bifurcation}
\label{sec:ell-hyp}

\begin{figure}[h]
  \centering
  \includegraphics[width=\linewidth]{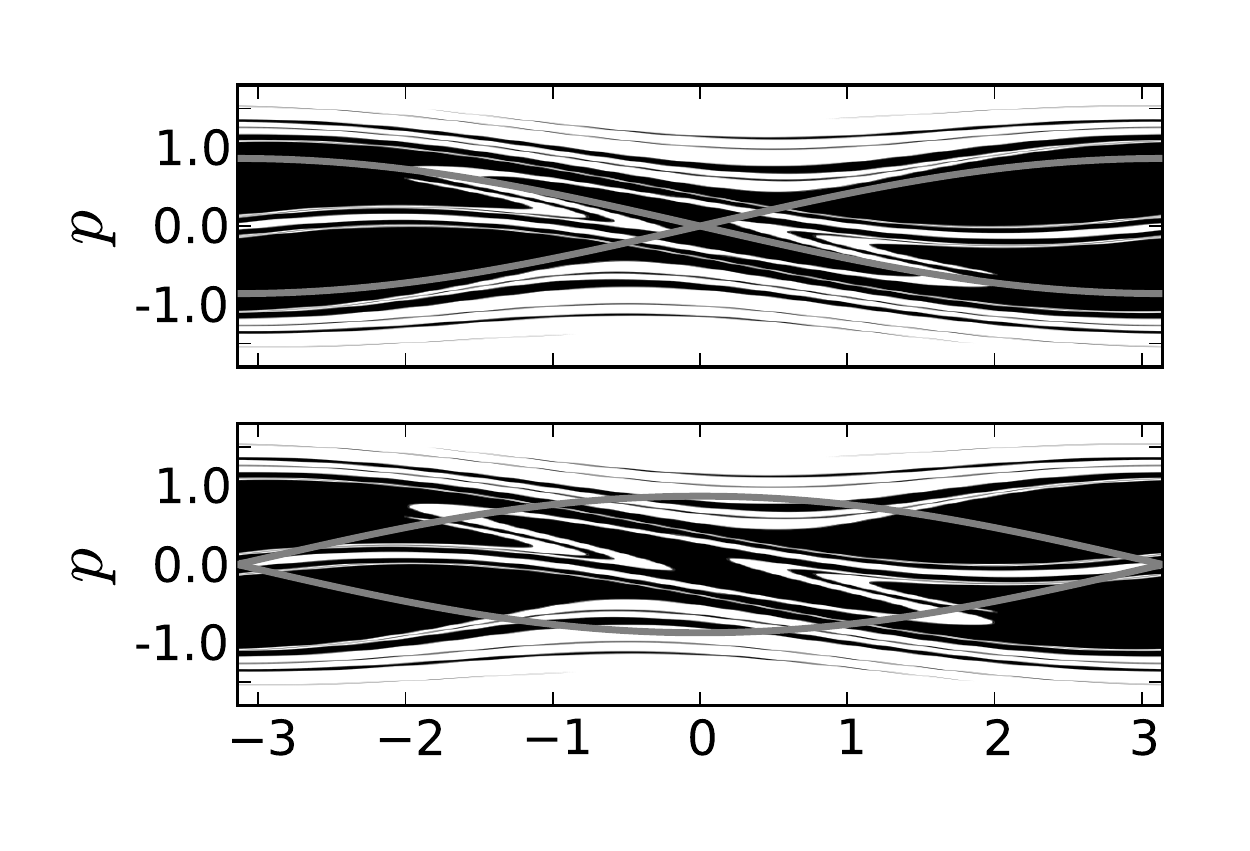}
  \caption{Phase space DF $f(\theta,p)$ for run i at time $t=20$ and $t=25$ (see Fig.~\ref{fig:moft-mixi}). Two counter-propagating clusters are found that lead to sign changes in $m_x$. The top panel displays a separatrix for $m_x<0$ and the bottom panel displays a separatrix for $m_x>0$.}
  \label{fig:ell-hyp}
\end{figure}

The study of run i, leading to the counter propagating resonances, displays an interesting bifurcation of the elliptic fixed point $(0,0)$. The nature of that point depends on the sign of $m_x$ that is found to change during time evolution.
The separatrices corresponding to the two situations of the elliptic-hyperbolic bifurcation are shown in Fig.~\ref{fig:ell-hyp}.

\begin{figure}[h!]
  \centering
  \includegraphics[width=\linewidth]{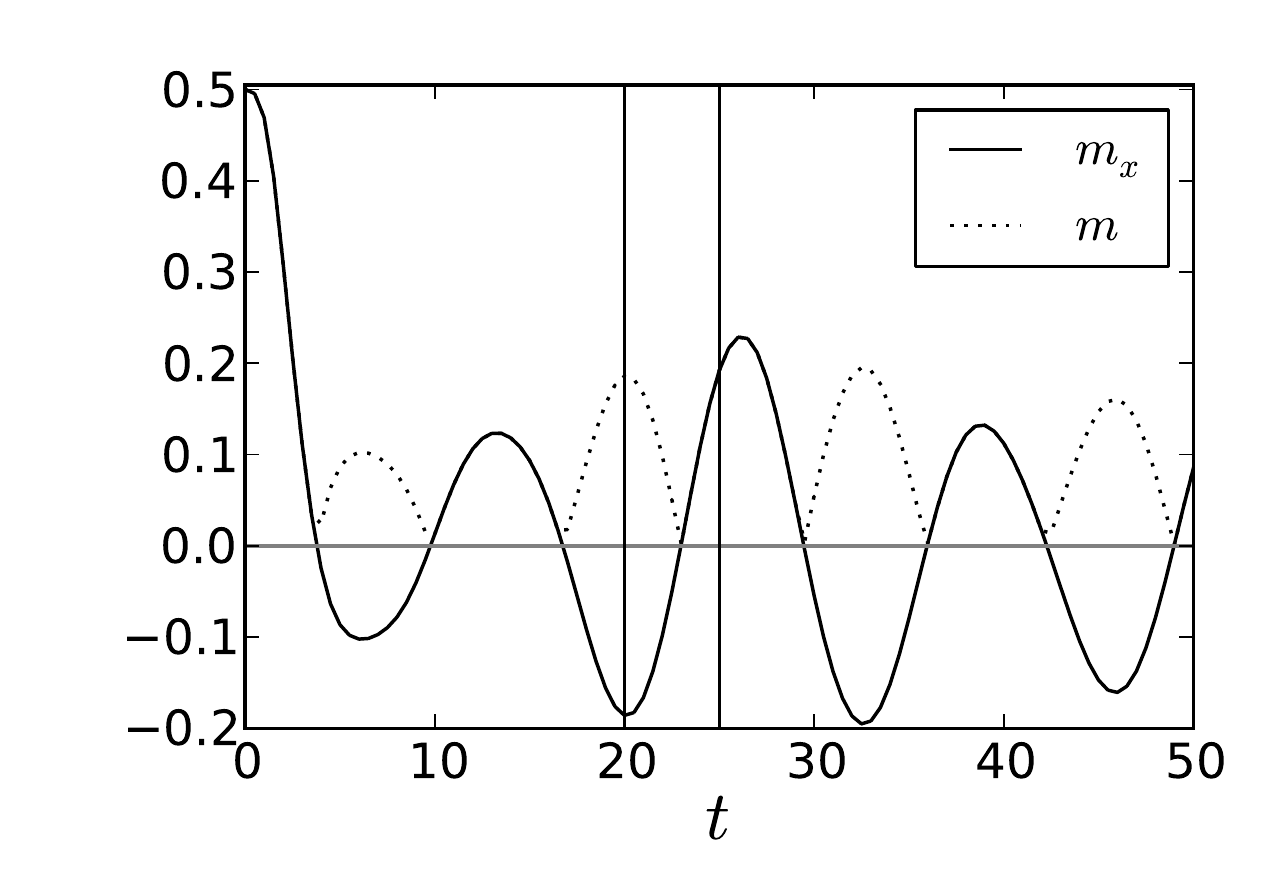}
  \caption{$m_x$ and $m$ as a function of time for run i. The vertical lines indicate the times at which the snapshots of Fig.~\ref{fig:ell-hyp} are taken.}
  \label{fig:moft-mixi}
\end{figure}

\begin{figure}[h]
  \centering
  \includegraphics[width=\linewidth]{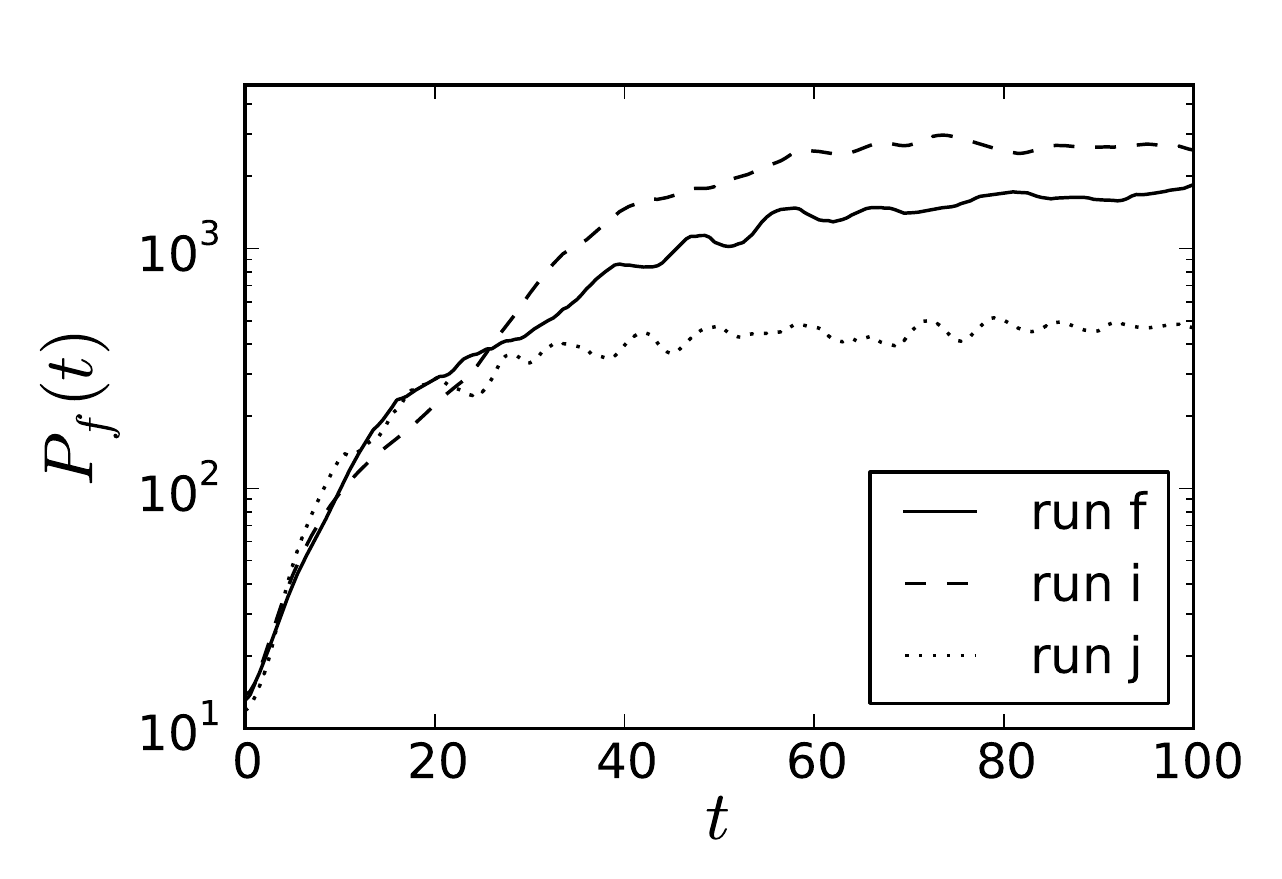}
  \caption{The perimeter $P_f(t)$ for the runs f, i and j of Table~\ref{tab:simus}. All runs are performed with the setting ``8kb'' of Table~\ref{tab:params}.}
  \label{fig:perim-all}
\end{figure}

The evolution of $m$ and $m_x$ are displayed in Fig.~\ref{fig:moft-mixi} ($m_y=0$ at all times).
Fluid elements that are located in the alternating separatrices region experience a strong amount of stretching and folding. This is confirmed in Fig.~\ref{fig:perim-all} where $P_f(t)$ is shown and compared to the result for run f.  The improved mixing in run~i should not however lead to premature conclusions about the relaxation of the systems on a macroscopic level. The counter propagating resonances are very stable and prevent the relaxation to complete equilibrium in which the resonances are not predicted.

In order to link the phenomenon of stretching and folding to the amount of evolution from an initial condition to a QSS, run j provides an interesting comparison.
The initial magnetization is $m_0=0.5$ and the magnetization predicted by Lynden-Bell's theory is also $m_{QSS}=0.5$. A large region of the initial waterbag will remain below the separatrix and smaller oscillations than for run f are expected. This is indeed found to be the case in Fig.~\ref{fig:perim-all}. A comparison between run f and run j for the numerical parameters 8kb is given in Fig.~\ref{fig:perim-all} confirming that the amount of stretching and folding needed for run j is well below the one for run f.

\section{Mean-field maps}
\label{sec:mfm}

In this section, we simplify the dynamics of the HMF model into discrete-time maps giving the mean-field approximation of symplectic coupled map systems \cite{KK92} ruled by the periodically kicked Hamiltonian:
\begin{eqnarray}
H &=& \sum_{i=1}^N \frac{p_i^2}{2} + \Big[ \sum_{i=1}^N U_0(x_i) \nonumber\\
&& +\frac{1}{2N} \sum_{i,j=1}^N U(x_i-x_j)\Big] \times \sum_{n=-\infty}^{+\infty} \delta (t-n) \, .
\end{eqnarray}
In the mean-field approximation, the positions and momenta of the particles after each kick are mapped by the time evolution according to
\begin{equation}
\Phi
\left\{
\begin{array}{l}
p_{n+1}=p_n-U'_0(x_n)-\int dx\, dp\, f(x,p) \,U'(x_n-x) \\
x_{n+1}=x_n+p_{n+1}\, ,
\end{array}
\right.
\end{equation}
where $n$ denotes the discrete time and $f(x,p)$ the distribution function representing
the ensemble of particles normalized by $\int f(x,p)\,dx\,dp=1$.  The map is area-preserving
so that the normalization condition is maintained during the time evolution.
Here, we consider the interaction $U(x)=C\left[1-\cos(2\pi x)\right]$ and the external
potentials: $U_0=Kx^2/2$ corresponding to Arnold's cat map \cite{PV87,S02} or $U_0=A \left[1-\cos(2\pi x)\right]/(2\pi)$ giving the standard map \cite{C79,LL83}.

We suppose that the distribution function $f(x,p)$ is uniform in a phase-space domain $D_n$, which evolves in time under the mean-field mapping $D_n=\Phi^n(D_0)$.  The area of this domain is preserved: ${\rm Area}(D_n)={\rm Area}(D_0)$.  The problem is now to determine the time evolution of the perimeter of the domain $D_n$.

\subsection{Mean-field cat map}

In this case, the map takes the following form:
\begin{equation}
\left\{
\begin{array}{l}
p_{n+1}=p_n+K x_n -m_x[f] \sin(2\pi x_n) + m_y[f] \cos(2\pi x_n)\\
x_{n+1}=x_n+p_{n+1} \qquad ({\rm modulo} \, 1) \, ,
\end{array}
\right.
\label{mfcm}
\end{equation}
with the mean field
\begin{equation}
{\bf m}[f]=m_x[f] + i \, m_y[f] = C \int dx\, dp \, f(x,p) \, e^{i 2\pi x}\, .
\label{meanfield}
\end{equation}
Moreover, we suppose that the initial distribution function $f(x,p)$ is uniform in a circle of radius $\frac{1}{4}$ centered on the point $x=p=\frac{1}{2}$ of the unit square, which defines the initial domain $D_0$.  After $n$ iterations, the distribution function is still uniform but in a domain $D_n$ resulting from the time evolution of the map (\ref{mfcm}).

\begin{figure}[h]
  \centering
  \includegraphics[width=\linewidth]{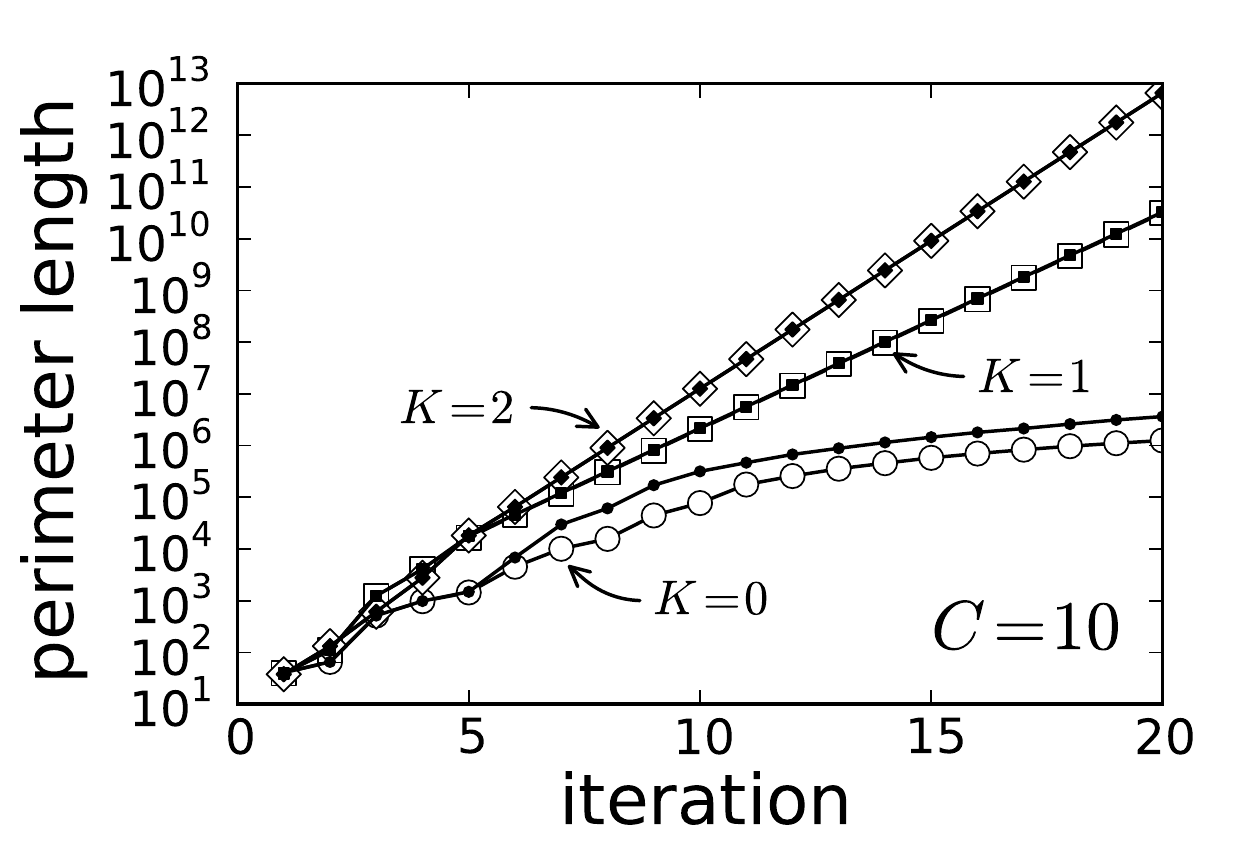}
  \caption{The perimeter $P_f(t)$ for the mean-field cat map with $K=0,1,2$ and $C=10$.
  The calculations are performed with $10^6$ (empty symbols) and $4\times 10^6$ (full symbols) points to represent the boundary and the bulk of the domain where the distribution function is uniform.}
  \label{fig10}
\end{figure}

If the coupling parameter vanishes, $C=0$, the mean field disappears and the map reduces to Arnold's cat map, which is known to be fully chaotic 
if the parameter $K$ is a positive integer \cite{PV87,S02}.
We notice that the hyperbolic character of the map persists as long as the coupling parameter $C$ is small enough that $K\gg 2\pi \vert {\bf m}\vert$.  Under this condition, we may expect that the hyperbolic character of the map will tend to stretch the domain $D_n$ into a filamentary phase-space structure, which produces an effective uniform distribution over the unit interval $x$ such that
the mean field (\ref{meanfield}) tends to vanish as $n\to\infty$.  For the same reason, the perimeter of the domain $D_n$ grows exponentially in time at a rate close to the Lyapunov exponent of the Arnold cat map.  This effect is indeed observed in Fig.\,\ref{fig10}, which depicts the perimeter of the domain $D_n$ versus the number $n$ of iterations for the parameter values $C=10$ and $K=0,1,2$.  The computation is performed for two approximations in which the boundary of the domain $D_n$ and its interior are discretized into $10^6$ and $4\times 10^6$ points.  In the hyperbolic regime for $K=1$ and $K=2$, the growth of the perimeter is exponential.  However, in the absence of chaos for $K=0$, the perimeter no longer increases exponentially and the computation is more sensitive to the discretization.  The fact is that for the value $C=10$ of the coupling parameter, the mean field vanishes so that an exponential stretching of the perimeter is not even induced by the mean field.

\subsection{Mean-field standard map}

Now, we consider the mean-field standard map defined on the unit square as
\begin{equation}
\left\{
\begin{array}{l}
p_{n+1}=p_n+ (A-m_x[f]) \sin(2\pi x_n) + m_y[f] \cos(2\pi x_n)\\
x_{n+1}=x_n+p_{n+1} \qquad ({\rm modulo} \, 1)\, ,
\end{array}
\right.
\label{mfstdm}
\end{equation}
with the same mean field as in Eq.\,(\ref{meanfield}).
Here, the single-particle potential as well as the one induced by the mean field
have a similar trigonometric form so that the mean field $m_x$ appears to have an effect comparable to the one of the constant parameter $A$ of the standard map 
(in the case where $m_y=0$) \cite{C79,LL83}.
The initial distribution function $f(x,p)$ is the same as in the previous subsection.

\begin{figure}[h]
  \centering
  \includegraphics[width=\linewidth]{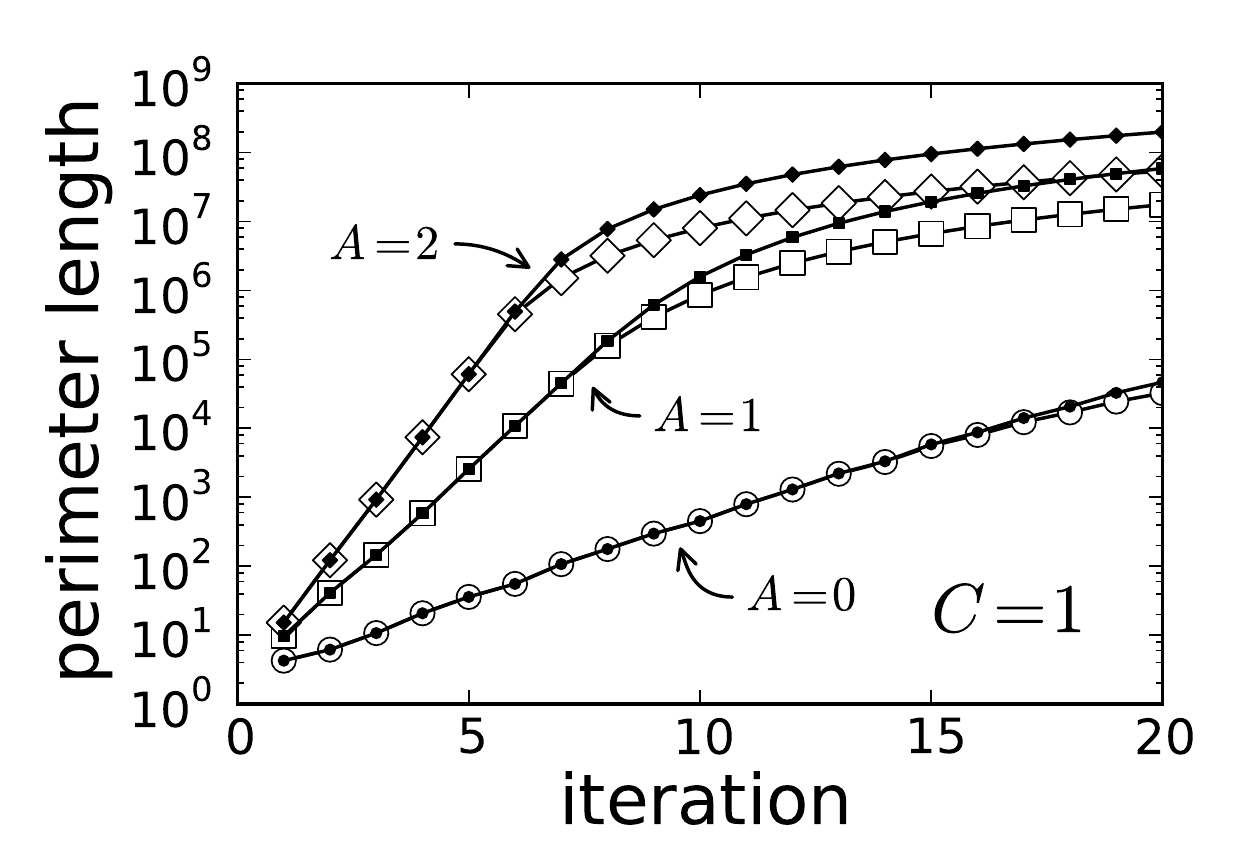}
  \caption{The perimeter $P_f(t)$ for the mean-field standard map with $K=0,1,2$ and $C=1$. The calculations are performed with $10^6$ (empty symbols) and $4\times 10^6$ (full symbols) points to represent the boundary and the bulk of the domain where the distribution function is uniform.}
  \label{fig11}
\end{figure}

Figure \ref{fig11} shows the growth of the perimeter of the domain $D_n$ as a function of the discrete time $n$ for the parameter values $C=1$ and $A=0,1,2$ for a discretization of the boundary and bulk of the domain $D_n$ into $10^6$ and $4\times 10^6$ points.  The early growth of the perimeter until about $n=6$ iterations is well approximated by these discretizations for $A=1,2$ and until about $n=15$ iterations for $A=0$.  Thereafter, both approximations deviate, showing the sensitivity of the calculation to the discretization of the domain $D_n$ and the distribution function $f(x,p)$ into points.  Nevertheless, the early growth appears to be exponential for $A=0,1,2$.

For $A=1,2$, the exponential increase of the perimeter can be explained by the global chaoticity of the standard map for $A>A_c\simeq 0.15$ \cite{C79,LL83}.  Indeed, the standard map is stretching domains at a rate equal to the Lyapunov exponent $\lambda\simeq \ln(\pi A)$, which is positive for $A=1,2$.  At each iteration, the perimeter of the domain $D_n$ is stretched by a factor $\exp(\lambda)>1$ for every point in the chaotic zones of the map.  The domain $D_n$ thus develops a filamentary phase-space structure uniformly distributed over the unit square so that the mean field (\ref{meanfield}) tends to vanish and the mean-field standard map (\ref{mfstdm}) behaves on long time as the single-particle standard map with $A=1,2$ and ${\bf m}=0$.  Accordingly, the exponential growth of the perimeter of the domain $D_n$ could go on as the discrete time $n$ increases.

In contrast, for $A=0$, the dynamics is entirely determined by the mean field (\ref{meanfield}).  It turns out that the mean field vanishes for $C>C_t$ with $C_t\simeq 2$, but is non-vanishing for $0<C<C_t$ by a phenomenon analogue to the one happening in the HMF model.  For $C=1$, the mean field fluctuates around the mean value $m_x\simeq -0.21$ with $m_y=0$.  This explains the early exponential growth of the perimeter observed in Fig.\,\ref{fig11} for $A=0$ and $C=1$.  Indeed, if we supposed that the mean field remains essentially constant, the Vlasov dynamics would be equivalent to the one of the simple standard map for $A=-m_x\simeq 0.21$.  In this case, the early growth rate would be about $\lambda\simeq 0.44$, which is close to the value $\lambda\simeq 0.48$ of the mean-field dynamics with $A=0$ and $C=1$ shown in Fig.\,\ref{fig11}. 

These results confirm that the stretching of the perimeter by the Vlasov dynamics is controlled by the local hyperbolicity of the effective single-particle phase-space dynamics induced by the current value of the mean field.  

\section{Conclusion}

Measuring the degree of mixing in kinetic theory is undoubtedly a considerable challenge.
However, in order to assess the evolution of a given system, a quantitative tool must be chosen. An analogy with fluid mechanics allowed us to consider the use of the perimeter of the fluid in phase space, viewed as a dyed region.

In both the Hamiltonian Mean-Field model and the mean-field cat and standard maps, our numerical experiments report an exponential growth of the perimeter, sign that exponential stretching occurs at least locally on the contour of the fluid. The fluid in phase space displays folding and stretching structures that are absent from similar non-interacting and non-chaotic systems.

The complex intertwining of $f=0$ and $f=f_0$ regions in phase space confirms that a coarse grained theory is able to provide a good description of the fluid after mixing has taken place.
The existence of an infinite number of conserved quantities in the Vlasov equation forbids the increase of entropic functionals, which implies that coarse graining is needed in order to describe the evolution towards stationary states (SS) or quasi-stationary states (QSS).
In the light of our findings about the mechanism leading to such QSS, violent relaxation can be understood from a pure Vlasovian picture in which the deformation of the initial condition allows the system to reach a QSS. This deformation takes place at constant phase-space volume but with a perimeter that increases exponentially during some lapses of the time evolution. The increase in the perimeter is made possible through successive stretching and folding the fluid.
Subsequently, finite $N$ effects may come into play but are of a different type and are not discussed herein.

The understanding of phenomena related to violent relaxation, for instance the separation of phase space in different regions~\cite{yamashiro_et_al_origin_of_core-halo} that leads to the formation of a core and a halo are coherent with the results of the present article.
Applying our method to more common systems, such as the self-gravitating sheet model, is expected to reveal a similar behavior and would prove interesting to complete the existing body of literature.

From our numerical observations, we have also found that stretching and folding cause a damping of oscillations (see Table~\ref{tab:std}) that competes with numerical oscillations.

As known from the literature, the presence of self-consistent resonances may organize phase space in a way that prevents the statistical prediction of stationary regimes. Indeed, these resonances imply that the energy distribution function is non monotonous, in contrast to the predictions of Lynden-Bell's theory or Boltzmann-Gibbs equilibrium. The self-consistent resonances are often at the origin of the so-called ``dynamical effects'' that prevent a system to reach a QSS or thermodynamical equilibrium.

Finally, let us recall that, in the perspective given in the present paper, the phase-space exponential separation leading to mixing has similarities to what happens in chaotic dynamics although notable differences exist such that the dependence on the initial distribution function as well as the time dependence of the stretching rate. A possible way to tighten both aspects is to use test particles that feel the time-dependent magnetization generated by a Vlasov simulation. It is possible to push this method further by the use of a Poincar{\'e} map on these test particles, providing one with the tools of iterative mapping, which are of common use in nonlinear dynamics.

\section*{Acknowledgments}

This research is financially supported by the Belgian Federal Government (Interuniversity Attraction Pole ``Nonlinear systems, stochastic processes, and statistical mechanics'', 2007-2011).


\begin{thebibliography}{42}
\expandafter\ifx\csname natexlab\endcsname\relax\def\natexlab#1{#1}\fi
\expandafter\ifx\csname bibnamefont\endcsname\relax
  \def\bibnamefont#1{#1}\fi
\expandafter\ifx\csname bibfnamefont\endcsname\relax
  \def\bibfnamefont#1{#1}\fi
\expandafter\ifx\csname citenamefont\endcsname\relax
  \def\citenamefont#1{#1}\fi
\expandafter\ifx\csname url\endcsname\relax
  \def\url#1{\texttt{#1}}\fi
\expandafter\ifx\csname urlprefix\endcsname\relax\def\urlprefix{URL }\fi
\providecommand{\bibinfo}[2]{#2}
\providecommand{\eprint}[2][]{\url{#2}}

\bibitem[{\citenamefont{Dauxois et~al.}(2002)\citenamefont{Dauxois, Ruffo,
  Arimondo, and Wilkens}}]{long-range-02}
\bibinfo{editor}{\bibfnamefont{T.}~\bibnamefont{Dauxois}},
  \bibinfo{editor}{\bibfnamefont{S.}~\bibnamefont{Ruffo}},
  \bibinfo{editor}{\bibfnamefont{E.}~\bibnamefont{Arimondo}}, \bibnamefont{and}
  \bibinfo{editor}{\bibfnamefont{M.}~\bibnamefont{Wilkens}}, eds.,
  \emph{\bibinfo{title}{Dynamics and Thermodynamics of Systems With Long Range
  Interactions}}, vol. \bibinfo{volume}{602} of \emph{\bibinfo{series}{Lecture
  Notes in Physics}} (\bibinfo{publisher}{Springer-Verlag},
  \bibinfo{year}{2002}).

\bibitem[{\citenamefont{Campa et~al.}(2008)\citenamefont{Campa, Giansanti,
  Morigi, and Sylos~Labini}}]{long-range-07}
\bibinfo{editor}{\bibfnamefont{A.}~\bibnamefont{Campa}},
  \bibinfo{editor}{\bibfnamefont{A.}~\bibnamefont{Giansanti}},
  \bibinfo{editor}{\bibfnamefont{G.}~\bibnamefont{Morigi}}, \bibnamefont{and}
  \bibinfo{editor}{\bibfnamefont{F.}~\bibnamefont{Sylos~Labini}}, eds.,
  \emph{\bibinfo{title}{Dynamics and Thermodynamics of Systems with Long Range
  Interactions: Theory and Experiments, Assisi, Italy 4-8 July 2007}}, vol.
  \bibinfo{volume}{970} of \emph{\bibinfo{series}{AIP Conference Proceedings}}
  (\bibinfo{publisher}{American Institute of Physics}, \bibinfo{year}{2008}).

\bibitem[{\citenamefont{Campa et~al.}(2009)\citenamefont{Campa, Dauxois, and
  Ruffo}}]{campa_et_al_phys_rep_2009}
\bibinfo{author}{\bibfnamefont{A.}~\bibnamefont{Campa}},
  \bibinfo{author}{\bibfnamefont{T.}~\bibnamefont{Dauxois}}, \bibnamefont{and}
  \bibinfo{author}{\bibfnamefont{S.}~\bibnamefont{Ruffo}},
  \bibinfo{journal}{Phys. Rep.} \textbf{\bibinfo{volume}{480}},
  \bibinfo{pages}{57} (\bibinfo{year}{2009}).

\bibitem[{\citenamefont{Yamaguchi et~al.}(2004)\citenamefont{Yamaguchi,
  Barr{\'e}, Bouchet, Dauxois, and Ruffo}}]{yamaguchi_et_al_physica_a_2004}
\bibinfo{author}{\bibfnamefont{Y.~Y.} \bibnamefont{Yamaguchi}},
  \bibinfo{author}{\bibfnamefont{J.}~\bibnamefont{Barr{\'e}}},
  \bibinfo{author}{\bibfnamefont{F.}~\bibnamefont{Bouchet}},
  \bibinfo{author}{\bibfnamefont{T.}~\bibnamefont{Dauxois}}, \bibnamefont{and}
  \bibinfo{author}{\bibfnamefont{S.}~\bibnamefont{Ruffo}},
  \bibinfo{journal}{Physica A} \textbf{\bibinfo{volume}{337}},
  \bibinfo{pages}{36} (\bibinfo{year}{2004}).

\bibitem[{\citenamefont{Bouchet and Dauxois}(2005)}]{bouchet_dauxois_pre_2005}
\bibinfo{author}{\bibfnamefont{F.}~\bibnamefont{Bouchet}} \bibnamefont{and}
  \bibinfo{author}{\bibfnamefont{T.}~\bibnamefont{Dauxois}},
  \bibinfo{journal}{Phys. Rev. E} \textbf{\bibinfo{volume}{72}},
  \bibinfo{pages}{45103} (\bibinfo{year}{2005}).

\bibitem[{\citenamefont{Jain et~al.}(2007)\citenamefont{Jain, Bouchet, and
  Mukamel}}]{jain_et_al_relaxation_times_2007}
\bibinfo{author}{\bibfnamefont{K.}~\bibnamefont{Jain}},
  \bibinfo{author}{\bibfnamefont{F.}~\bibnamefont{Bouchet}}, \bibnamefont{and}
  \bibinfo{author}{\bibfnamefont{D.}~\bibnamefont{Mukamel}},
  \bibinfo{journal}{J. Stat. Mech.} \textbf{\bibinfo{volume}{11}},
  \bibinfo{pages}{11008} (\bibinfo{year}{2007}).

\bibitem[{\citenamefont{Latora et~al.}(1999)\citenamefont{Latora, Rapisarda,
  and Ruffo}}]{latora_et_al_prl_1999}
\bibinfo{author}{\bibfnamefont{V.}~\bibnamefont{Latora}},
  \bibinfo{author}{\bibfnamefont{A.}~\bibnamefont{Rapisarda}},
  \bibnamefont{and} \bibinfo{author}{\bibfnamefont{S.}~\bibnamefont{Ruffo}},
  \bibinfo{journal}{Phys. Rev. Lett.} \textbf{\bibinfo{volume}{83}},
  \bibinfo{pages}{2104} (\bibinfo{year}{1999}).

\bibitem[{\citenamefont{Barr{\'e} et~al.}(2004)\citenamefont{Barr{\'e},
  Dauxois, {De Ninno}, Fanelli, and Ruffo}}]{barre_et_al_pre_2004}
\bibinfo{author}{\bibfnamefont{J.}~\bibnamefont{Barr{\'e}}},
  \bibinfo{author}{\bibfnamefont{T.}~\bibnamefont{Dauxois}},
  \bibinfo{author}{\bibfnamefont{G.}~\bibnamefont{{De Ninno}}},
  \bibinfo{author}{\bibfnamefont{D.}~\bibnamefont{Fanelli}}, \bibnamefont{and}
  \bibinfo{author}{\bibfnamefont{S.}~\bibnamefont{Ruffo}},
  \bibinfo{journal}{Phys. Rev. E} \textbf{\bibinfo{volume}{69}},
  \bibinfo{pages}{045501} (\bibinfo{year}{2004}).

\bibitem[{\citenamefont{Yamaguchi}(2008)}]{yamaguchi_pre_2008}
\bibinfo{author}{\bibfnamefont{Y.~Y.} \bibnamefont{Yamaguchi}},
  \bibinfo{journal}{Phys. Rev. E} \textbf{\bibinfo{volume}{78}},
  \bibinfo{pages}{041114} (\bibinfo{year}{2008}).

\bibitem[{\citenamefont{Bachelard et~al.}(2010)\citenamefont{Bachelard, Manos,
  de~Buyl, Staniscia, Cataliotti, Ninno, Fanelli, and
  Piovella}}]{bachelard_et_al_jstat_2010}
\bibinfo{author}{\bibfnamefont{R.}~\bibnamefont{Bachelard}},
  \bibinfo{author}{\bibfnamefont{T.}~\bibnamefont{Manos}},
  \bibinfo{author}{\bibfnamefont{P.}~\bibnamefont{de~Buyl}},
  \bibinfo{author}{\bibfnamefont{F.}~\bibnamefont{Staniscia}},
  \bibinfo{author}{\bibfnamefont{F.~S.} \bibnamefont{Cataliotti}},
  \bibinfo{author}{\bibfnamefont{G.~D.} \bibnamefont{Ninno}},
  \bibinfo{author}{\bibfnamefont{D.}~\bibnamefont{Fanelli}}, \bibnamefont{and}
  \bibinfo{author}{\bibfnamefont{N.}~\bibnamefont{Piovella}},
  \bibinfo{journal}{J. Stat. Mech.} \textbf{\bibinfo{volume}{2010}},
  \bibinfo{pages}{P06009} (\bibinfo{year}{2010}).

\bibitem[{\citenamefont{H{\'e}non}(1964)}]{henon_1964}
\bibinfo{author}{\bibfnamefont{M.}~\bibnamefont{H{\'e}non}},
  \bibinfo{journal}{Annales d'Astrophysique} \textbf{\bibinfo{volume}{27}},
  \bibinfo{pages}{83} (\bibinfo{year}{1964}).

\bibitem[{\citenamefont{Lecar}(1966)}]{lecar_iaus_1966}
\bibinfo{author}{\bibfnamefont{M.}~\bibnamefont{Lecar}}, \bibinfo{journal}{The
  Theory of Orbits in the Solar System and in Stellar Systems. Proceedings from
  Symposium no. 25 held in Thessaloniki} \textbf{\bibinfo{volume}{25}},
  \bibinfo{pages}{46} (\bibinfo{year}{1966}).

\bibitem[{\citenamefont{Lynden-Bell}(1967)}]{lynden-bell_1967}
\bibinfo{author}{\bibfnamefont{D.}~\bibnamefont{Lynden-Bell}},
  \bibinfo{journal}{Mon. Not. R. Astron. Soc.} \textbf{\bibinfo{volume}{136}},
  \bibinfo{pages}{101} (\bibinfo{year}{1967}).

\bibitem[{\citenamefont{Antoniazzi
  et~al.}(2007{\natexlab{a}})\citenamefont{Antoniazzi, Fanelli, Ruffo, and
  Yamaguchi}}]{antoniazzi_et_al_prl_2007}
\bibinfo{author}{\bibfnamefont{A.}~\bibnamefont{Antoniazzi}},
  \bibinfo{author}{\bibfnamefont{D.}~\bibnamefont{Fanelli}},
  \bibinfo{author}{\bibfnamefont{S.}~\bibnamefont{Ruffo}}, \bibnamefont{and}
  \bibinfo{author}{\bibfnamefont{Y.~Y.} \bibnamefont{Yamaguchi}},
  \bibinfo{journal}{Phys. Rev. Lett.} \textbf{\bibinfo{volume}{99}},
  \bibinfo{pages}{040601} (\bibinfo{year}{2007}{\natexlab{a}}).

\bibitem[{\citenamefont{Hohl}(1968)}]{hohl_NASA_1968}
\bibinfo{author}{\bibfnamefont{F.}~\bibnamefont{Hohl}},
  \emph{\bibinfo{title}{Theory and results on collective and collisional
  effects for a one-dimensional self-gravitating system}},
  \bibinfo{howpublished}{NASA Technical Report R--289} (\bibinfo{year}{1968}).

\bibitem[{\citenamefont{Luwel and Severne}(1985)}]{luwel_severne_1985}
\bibinfo{author}{\bibfnamefont{M.}~\bibnamefont{Luwel}} \bibnamefont{and}
  \bibinfo{author}{\bibfnamefont{G.}~\bibnamefont{Severne}},
  \bibinfo{journal}{{A}stron. {A}strophys.} \textbf{\bibinfo{volume}{152}},
  \bibinfo{pages}{305} (\bibinfo{year}{1985}).

\bibitem[{\citenamefont{Mineau et~al.}(1990)\citenamefont{Mineau, Feix, and
  Rouet}}]{mineau_et_al_numerical_holes_1990}
\bibinfo{author}{\bibfnamefont{P.}~\bibnamefont{Mineau}},
  \bibinfo{author}{\bibfnamefont{M.~R.} \bibnamefont{Feix}}, \bibnamefont{and}
  \bibinfo{author}{\bibfnamefont{J.~L.} \bibnamefont{Rouet}},
  \bibinfo{journal}{{A}stron. {A}strophys.} \textbf{\bibinfo{volume}{228}},
  \bibinfo{pages}{344} (\bibinfo{year}{1990}).

\bibitem[{\citenamefont{Funato et~al.}(1992)\citenamefont{Funato, Makino, and
  Ebisuzaki}}]{funato_et_al_not_relaxation_1992}
\bibinfo{author}{\bibfnamefont{Y.}~\bibnamefont{Funato}},
  \bibinfo{author}{\bibfnamefont{J.}~\bibnamefont{Makino}}, \bibnamefont{and}
  \bibinfo{author}{\bibfnamefont{T.}~\bibnamefont{Ebisuzaki}},
  \bibinfo{journal}{Publ. Astron. Soc. Japan} \textbf{\bibinfo{volume}{44}},
  \bibinfo{pages}{613} (\bibinfo{year}{1992}).

\bibitem[{\citenamefont{Levin et~al.}(2008)\citenamefont{Levin, Pakter, and
  Teles}}]{levin_et_al_plasmas_prl_2008}
\bibinfo{author}{\bibfnamefont{Y.}~\bibnamefont{Levin}},
  \bibinfo{author}{\bibfnamefont{R.}~\bibnamefont{Pakter}}, \bibnamefont{and}
  \bibinfo{author}{\bibfnamefont{T.~N.} \bibnamefont{Teles}},
  \bibinfo{journal}{Phys. Rev. Lett.} \textbf{\bibinfo{volume}{100}},
  \bibinfo{pages}{40604} (\bibinfo{year}{2008}).

\bibitem[{\citenamefont{Teles et~al.}(2010)\citenamefont{Teles, Levin, Pakter,
  and Rizzato}}]{teles_et_al_2d_self-grav_2010}
\bibinfo{author}{\bibfnamefont{T.}~\bibnamefont{Teles}},
  \bibinfo{author}{\bibfnamefont{Y.}~\bibnamefont{Levin}},
  \bibinfo{author}{\bibfnamefont{R.}~\bibnamefont{Pakter}}, \bibnamefont{and}
  \bibinfo{author}{\bibfnamefont{F.}~\bibnamefont{Rizzato}},
  \bibinfo{journal}{J. Stat. Mech.} \textbf{\bibinfo{volume}{2010}},
  \bibinfo{pages}{P05007} (\bibinfo{year}{2010}).

\bibitem[{\citenamefont{Joyce and
  Worrakitpoonpon}(2010)}]{joyce_worrakitpoonpon_jstat_2010}
\bibinfo{author}{\bibfnamefont{M.}~\bibnamefont{Joyce}} \bibnamefont{and}
  \bibinfo{author}{\bibfnamefont{T.}~\bibnamefont{Worrakitpoonpon}},
  \bibinfo{journal}{J. Stat. Mech.} \textbf{\bibinfo{volume}{2010}},
  \bibinfo{pages}{P10012} (\bibinfo{year}{2010}).

\bibitem[{\citenamefont{Joyce and
  Worrakitpoonpon}(2011)}]{joyce_worrakitpoonpon_pre_2011}
\bibinfo{author}{\bibfnamefont{M.}~\bibnamefont{Joyce}} \bibnamefont{and}
  \bibinfo{author}{\bibfnamefont{T.}~\bibnamefont{Worrakitpoonpon}},
  \bibinfo{journal}{Phys. Rev. E} \textbf{\bibinfo{volume}{84}},
  \bibinfo{pages}{011139} (\bibinfo{year}{2011}).

\bibitem[{\citenamefont{Balescu}(1997)}]{balescu_statistical_dynamics}
\bibinfo{author}{\bibfnamefont{R.}~\bibnamefont{Balescu}},
  \emph{\bibinfo{title}{Statistical {D}ynamics - {M}atter out of
  {E}quilibrium}} (\bibinfo{publisher}{Imperial College Press},
  \bibinfo{address}{Imperial College -- London}, \bibinfo{year}{1997}).

\bibitem[{\citenamefont{O'neil and Coroniti}(1999)}]{oneil_coroniti_rmp_1999}
\bibinfo{author}{\bibfnamefont{T.}~\bibnamefont{O'neil}} \bibnamefont{and}
  \bibinfo{author}{\bibfnamefont{F.}~\bibnamefont{Coroniti}},
  \bibinfo{journal}{Rev. Mod. Phys.} \textbf{\bibinfo{volume}{71}},
  \bibinfo{pages}{S404} (\bibinfo{year}{1999}).

\bibitem[{\citenamefont{Antoni and Ruffo}(1995)}]{antoni_ruffo_1995}
\bibinfo{author}{\bibfnamefont{M.}~\bibnamefont{Antoni}} \bibnamefont{and}
  \bibinfo{author}{\bibfnamefont{S.}~\bibnamefont{Ruffo}},
  \bibinfo{journal}{Phys. Rev. E} \textbf{\bibinfo{volume}{52}},
  \bibinfo{pages}{2361} (\bibinfo{year}{1995}).

\bibitem[{\citenamefont{Elskens and
  Escande}(1991)}]{elskens_escande_nonlinearity_1991}
\bibinfo{author}{\bibfnamefont{Y.}~\bibnamefont{Elskens}} \bibnamefont{and}
  \bibinfo{author}{\bibfnamefont{D.~F.} \bibnamefont{Escande}},
  \bibinfo{journal}{Nonlinearity} \textbf{\bibinfo{volume}{4}},
  \bibinfo{pages}{615} (\bibinfo{year}{1991}).

\bibitem[{\citenamefont{Percival and Vivaldi}(1987)}]{PV87}
\bibinfo{author}{\bibfnamefont{I.}~\bibnamefont{Percival}} \bibnamefont{and}
  \bibinfo{author}{\bibfnamefont{F.}~\bibnamefont{Vivaldi}},
  \bibinfo{journal}{Physica D} \textbf{\bibinfo{volume}{27}},
  \bibinfo{pages}{373} (\bibinfo{year}{1987}).

\bibitem[{\citenamefont{Sano}(2002)}]{S02}
\bibinfo{author}{\bibfnamefont{M.~M.} \bibnamefont{Sano}},
  \bibinfo{journal}{Phys. Rev. E} \textbf{\bibinfo{volume}{66}},
  \bibinfo{pages}{046211} (\bibinfo{year}{2002}).

\bibitem[{\citenamefont{Chirikov}(1979)}]{C79}
\bibinfo{author}{\bibfnamefont{B.~V.} \bibnamefont{Chirikov}},
  \bibinfo{journal}{Phys. Rep.} \textbf{\bibinfo{volume}{52}},
  \bibinfo{pages}{265} (\bibinfo{year}{1979}).

\bibitem[{\citenamefont{Lichtenberg and Lieberman}(1983)}]{LL83}
\bibinfo{author}{\bibfnamefont{A.~J.} \bibnamefont{Lichtenberg}}
  \bibnamefont{and} \bibinfo{author}{\bibfnamefont{M.~A.}
  \bibnamefont{Lieberman}}, \emph{\bibinfo{title}{{R}egular and {S}tochastic
  {M}otion}} (\bibinfo{publisher}{Springer, New York}, \bibinfo{year}{1983}).

\bibitem[{\citenamefont{Antoniazzi
  et~al.}(2007{\natexlab{b}})\citenamefont{Antoniazzi, Califano, Fanelli, and
  Ruffo}}]{antoniazzi_califano_prl}
\bibinfo{author}{\bibfnamefont{A.}~\bibnamefont{Antoniazzi}},
  \bibinfo{author}{\bibfnamefont{F.}~\bibnamefont{Califano}},
  \bibinfo{author}{\bibfnamefont{D.}~\bibnamefont{Fanelli}}, \bibnamefont{and}
  \bibinfo{author}{\bibfnamefont{S.}~\bibnamefont{Ruffo}},
  \bibinfo{journal}{Phys. Rev. Lett.} \textbf{\bibinfo{volume}{98}},
  \bibinfo{pages}{150602} (\bibinfo{year}{2007}{\natexlab{b}}).

\bibitem[{\citenamefont{Sonnendrucker et~al.}(1999)\citenamefont{Sonnendrucker,
  Roche, Bertrand, and Ghizzo}}]{sonnendrucker_et_al_semi-lag_1999}
\bibinfo{author}{\bibfnamefont{E.}~\bibnamefont{Sonnendrucker}},
  \bibinfo{author}{\bibfnamefont{J.}~\bibnamefont{Roche}},
  \bibinfo{author}{\bibfnamefont{P.}~\bibnamefont{Bertrand}}, \bibnamefont{and}
  \bibinfo{author}{\bibfnamefont{A.}~\bibnamefont{Ghizzo}},
  \bibinfo{journal}{J. Comp. Phys.} \textbf{\bibinfo{volume}{149}},
  \bibinfo{pages}{201} (\bibinfo{year}{1999}).

\bibitem[{\citenamefont{de~Buyl}(2010)}]{de_buyl_cnsns_2010}
\bibinfo{author}{\bibfnamefont{P.}~\bibnamefont{de~Buyl}},
  \bibinfo{journal}{Commun. Nonlinear Sci. Numer. Simulat.}
  \textbf{\bibinfo{volume}{15}}, \bibinfo{pages}{2133} (\bibinfo{year}{2010}).

\bibitem[{\citenamefont{Califano et~al.}(2006)\citenamefont{Califano, Galeotti,
  and Mangeney}}]{califano_galeotti_pop_2006}
\bibinfo{author}{\bibfnamefont{F.}~\bibnamefont{Califano}},
  \bibinfo{author}{\bibfnamefont{L.}~\bibnamefont{Galeotti}}, \bibnamefont{and}
  \bibinfo{author}{\bibfnamefont{A.}~\bibnamefont{Mangeney}},
  \bibinfo{journal}{Phys. Plasmas} \textbf{\bibinfo{volume}{13}},
  \bibinfo{pages}{082102} (\bibinfo{year}{2006}).

\bibitem[{\citenamefont{de~Buyl et~al.}(2011)\citenamefont{de~Buyl, Mukamel,
  and Ruffo}}]{de_buyl_et_al_rsta_2011}
\bibinfo{author}{\bibfnamefont{P.}~\bibnamefont{de~Buyl}},
  \bibinfo{author}{\bibfnamefont{D.}~\bibnamefont{Mukamel}}, \bibnamefont{and}
  \bibinfo{author}{\bibfnamefont{S.}~\bibnamefont{Ruffo}},
  \bibinfo{journal}{Phil. Trans. R. Soc. A} \textbf{\bibinfo{volume}{369}},
  \bibinfo{pages}{439} (\bibinfo{year}{2011}).

\bibitem[{\citenamefont{de~Buyl et~al.}(2010)\citenamefont{de~Buyl, Mukamel,
  and Ruffo}}]{de_buyl_et_al_in_prep}
\bibinfo{author}{\bibfnamefont{P.}~\bibnamefont{de~Buyl}},
  \bibinfo{author}{\bibfnamefont{D.}~\bibnamefont{Mukamel}}, \bibnamefont{and}
  \bibinfo{author}{\bibfnamefont{S.}~\bibnamefont{Ruffo}},
  \bibinfo{howpublished}{arXiv:1012.2594} (\bibinfo{year}{2010}).

\bibitem[{\citenamefont{Ottino}(1989)}]{ottino_book_1989}
\bibinfo{author}{\bibfnamefont{J.~M.} \bibnamefont{Ottino}},
  \emph{\bibinfo{title}{The kinematics of mixing: stretching, chaos and
  transport}} (\bibinfo{publisher}{Cambridge University Press},
  \bibinfo{address}{Cambridge, England}, \bibinfo{year}{1989}).

\bibitem[{\citenamefont{del Castillo-Negrete and
  Firpo}(2002)}]{del-castillo-negrete_firpo_chaos_2002}
\bibinfo{author}{\bibfnamefont{D.}~\bibnamefont{del Castillo-Negrete}}
  \bibnamefont{and} \bibinfo{author}{\bibfnamefont{M.-C.} \bibnamefont{Firpo}},
  \bibinfo{journal}{Chaos} \textbf{\bibinfo{volume}{12}}, \bibinfo{pages}{496}
  (\bibinfo{year}{2002}).

\bibitem[{\citenamefont{Wiggins and Ottino}(2004)}]{wiggins_ottino_rsta_2004}
\bibinfo{author}{\bibfnamefont{S.}~\bibnamefont{Wiggins}} \bibnamefont{and}
  \bibinfo{author}{\bibfnamefont{J.~M.} \bibnamefont{Ottino}},
  \bibinfo{journal}{Phil. Trans. R. Soc. A} \textbf{\bibinfo{volume}{362}},
  \bibinfo{pages}{937} (\bibinfo{year}{2004}).

\bibitem[{\citenamefont{Bourke}(1987)}]{bourke_conrec_1987}
\bibinfo{author}{\bibfnamefont{P.}~\bibnamefont{Bourke}},
  \bibinfo{journal}{Byte Magazine}  (\bibinfo{year}{1987}).

\bibitem[{\citenamefont{Konishi and Kaneko}(1992)}]{KK92}
\bibinfo{author}{\bibfnamefont{T.}~\bibnamefont{Konishi}} \bibnamefont{and}
  \bibinfo{author}{\bibfnamefont{K.}~\bibnamefont{Kaneko}},
  \bibinfo{journal}{J. Phys. A: Math. Gen.} \textbf{\bibinfo{volume}{25}},
  \bibinfo{pages}{6283} (\bibinfo{year}{1992}).

\bibitem[{\citenamefont{Yamashiro et~al.}(1992)\citenamefont{Yamashiro, Gouda,
  and Sakagami}}]{yamashiro_et_al_origin_of_core-halo}
\bibinfo{author}{\bibfnamefont{T.}~\bibnamefont{Yamashiro}},
  \bibinfo{author}{\bibfnamefont{N.}~\bibnamefont{Gouda}}, \bibnamefont{and}
  \bibinfo{author}{\bibfnamefont{M.}~\bibnamefont{Sakagami}},
  \bibinfo{journal}{Prog. Theor. Phys.} \textbf{\bibinfo{volume}{88}},
  \bibinfo{pages}{269} (\bibinfo{year}{1992}).

\end{thebibliography}
\end{document}